# A spin-filter for polarized electron acceleration in plasma wakefields

Yitong Wu[1,2], Liangliang Ji[1,3*], Xuesong Geng[1,2], Johannes Thomas[4], Markus Büscher[5,6], Alexander Pukhov[4], Anna Hützen[5,6], Lingang Zhang[1], Baifei Shen[1,3,7§], and Ruxin Li[1,3,8†]

[1]*State Key Laboratory of High Field Laser Physics, Shanghai Institute of Optics and Fine Mechanics, Chinese Academy of Sciences, Shanghai 201800, China*

[2]*Center of Materials Science and Optoelectronics Engineering, University of Chinese Academy of Sciences, Beijing 100049, China*

[3]*CAS Center for Excellence in Ultra-intense Laser Science, Shanghai 201800, China*

[4]*Institut für Theoretische Physik I, Heinrich-Heine-Universität Düsseldorf, Universitätsstr. 1, 40225 Düsseldorf, Germany*

[5]*Peter Grünberg Institut (PGI-6), Forschungszentrum Jülich, Wilhelm-Johnen-Str. 1, 52425 Jülich, Germany*

[6]*Institut für Laser- und Plasmaphysik, Heinrich-Heine-Universität Düsseldorf, Universitätsstr. 1, 40225 Düsseldorf, Germany*

[7]*Shanghai Normal University, Shanghai 200234, China*

[8]*Shanghai Tech University, Shanghai 201210, China*

## Abstract

We propose a filter method to generate electron beams of high polarization from bubble and blow-out wakefield accelerators. The mechanism is based on the idea to identify all electron-beam subsets with low-polarization and to filter them out by an X-shaped slit placed right behind the plasma accelerator. To find these subsets we investigate the dependence between the initial azimuthal angle and the spin of single electrons during the trapping process. This dependence shows that transverse electron spins preserve their orientation during injection if they are initially aligned parallel or anti-parallel to the local magnetic field. We derive a precise correlation of the local beam polarization as a function of the coordinate and the electron phase angle. Three-dimensional particle-in-cell simulations, incorporating classical spin dynamics, show that the beam polarization can be increased from 35% to about 80% after spin filtering. The injected flux is strongly restricted to preserve the beam polarization, e.g. <1kA in Ref. [27]. This limitation is removed by employing the proposed filter mechanism. The robust of the method is discussed that contains drive beam fluctuations, jitters, the thickness of the filter and initial temperature. This idea marks an efficient and simple strategy to generate energetic polarized electron beams based on wakefield acceleration.

# I. INTRODUCTION

Wakefield acceleration driven by either laser pulses (LWFA) [1] or electron beams (PWFA) [2] is advantageous due to its large acceleration gradient, reaching near GeV/cm [3] which is about over 3 orders of magnitudes higher than traditional accelerators [4]. Generation of GeV-level electron beams from both LWFA [5-7] and PWFA [8-10] have been realized recently, among which 8 GeV electron energy is obtained via LWFA [6] and several GeV via PWFA [9, 10]. These results pave a path to facilitate table-top accelerators and secondary light sources. An ultimate goal of wakefield acceleration is the pursuit of future electron-positron colliders [11,12]. Particularly, the electron and positron beams are favorable to be spin-polarized for the colliders. There are at least three main advantages for applying polarized beams: (i) The cross section is enhanced for certain reaction processes [13-16]. (ii) Unwanted background processes and reaction channels can be suppressed with appropriate combinations of the electron and positron beam polarizations [14,15]. (iii) By choosing suitable observables in the scattering processes, additional information such as quantum numbers and chiral couplings can be obtained [14,16]. Furthermore, energetic polarized electron beams are also used to generate polarized photons [17] and positrons [18] as well as to study material science and nuclear physics [19-21].

Conventionally, high energy polarized electron beams are obtained either from storage rings [19,20] based on radiative polarization (the Sokolov-Ternov effect [22], similar effects using ultra intense lasers are shown in Ref. [23]) or by extracting polarized electrons directly (via photocathodes [21,24], spin filters [21,25] or beam splitters [21,26]) for subsequent acceleration in Linacs. Recently polarized electron generation based on plasma wakefields has been explored in simulations [27-29]. The idea includes preparation of pre-polarized targets via the photodissociation of hydrogen halides [30-34]. The pre-polarized electrons suffer from depolarization in plasma wakefield due to the spin precession in self-generated magnetic field associated with the injected electrons. In order to preserve the electron polarization during the entire acceleration process, strong limitations must be imposed on the beam parameters, the plasma density as well as the injected beam flux. This can be mitigated by either restricting the electron flux to <1kA at laser amplitudes $a$<1.1 as proposed in Ref. [27] or employing vortex wakefield structures [28].

In this paper, we proposed a filter mechanism by selecting targeted electrons out of the whole accelerated beam to achieve polarization>80%. The mechanism largely removes the flux limitation issue on injected electrons. It relies on the fact that the beam polarization is dependent on the spatial/momentum angles-a universal phenomenon in density down-ramp injection. Specifically, in PIC simulations incorporating spin dynamics we show that spin precession in wakefild is significantly suppressed when the initial spin is parallel or anti-parallel to the local azimuthal magnetic field, leading to beam polarization dependence on the transverse coordinate angle in the plasma bubble. We then propose to select electrons from the related region in the transverse phase space with an X-shaped slit, which successfully maximizes the polarization of the accelerated beam. This unique feature has not been revealed by any previous studies and applies to transverse polarization.

## II. SIMULATION METHODS AND SETUPS

In order to demonstrate the spin-filter mechanism, we carry out full three-dimension (3D) particle-in-cell (PIC) simulations with the code VLPL (Virtual Laser Plasma Lab) [35]. To investigate spin dynamics, we integrate the spin precession into PIC code following the Thomas-Bargmann-Michel-Telegdi (T-BMT) equation [36]:

$$d\boldsymbol{s}/dt = \boldsymbol{\Omega} \times \boldsymbol{s} \quad (1a)$$

$$\boldsymbol{\Omega} = \frac{e}{m}\left(\frac{1}{\gamma}\boldsymbol{B} - \frac{\boldsymbol{\beta}}{\gamma+1}\times\frac{\boldsymbol{E}}{c}\right) + a_e\frac{e}{m}\left(\boldsymbol{B} - \frac{\gamma}{\gamma+1}\boldsymbol{\beta}(\boldsymbol{\beta}\cdot\boldsymbol{B}) - \boldsymbol{\beta}\times\frac{\boldsymbol{E}}{c}\right) \quad (1b)$$

Here $e$ is the elementary charge, $m$ the electron mass, $\boldsymbol{v}$ the electron velocity, and $\gamma=1/(1-v^2/c^2)^{1/2}$ the relativistic factor. $\boldsymbol{s}$ represents the normalized particle spin-direction vector with $|\boldsymbol{s}|=1$ following Ehrenfest's theorem [19]. $a_e=(g-2)/2\approx1.16\times10^{-3}$ (gyromagnetic factor $g$), while the vector $\boldsymbol{\Omega}$ is the precession frequency. As mentioned in Refs. [27-29], other effects such as the Stern-Gerlach force (S-G, gradient force), the Sokolov-Ternov effect (S-T, radiative polarization) [22] and electrostatic Coulomb collisions [37-38] are negligibly small for wakefield acceleration. Comparing the S-G force to the Lorentz force of wakefield acceleration $|F_{SG}/F_L|\sim|\nabla(\boldsymbol{S}\cdot\boldsymbol{B})/\gamma_e^2 cBm_e|\sim\hbar/\lambda m_e c\gamma_e^2\sim10^{-7}$ [19,28] suggests the drift motion caused by such a force is small enough to be neglected. With regard to the S-T effect, the typical polarization time is about $T_{pol,S-T}=8m_e^5c^8/5\sqrt{3}\hbar e^5F^3\gamma_e^2$ [19,28]. For typical wakefield acceleration ($\gamma_e\sim10^3$ and field strength F$\sim10^{16}$V/m), one has $T_{pol,S-T}\sim1\mu s$, corresponding to about 300m acceleration distances, which is much larger than typical wakefield acceleration distance.

In all the simulations, in order to ensure high injection efficiency [39-41], a transversely pre-polarized target (along the +$z$ axis) with a density bump is assumed. According to previous works [27-29], such a plasma density is usually parametrized by a ramp-shaped profile, $n(\kappa)=\{[\alpha-\Theta(\kappa)]\Theta(L-$

$\kappa)\cos(\pi\kappa/2L)+\Theta(\kappa-L)\}n_0$, where $\Theta(x)$ is the step function, $\kappa=x-x_p$, $x_p=36\mu m$, $L=16$ μm, $n_0=10^{18}cm^{-3}$. $\alpha=n_p/n_0=4$ is the ratio between the peak density of the ramp and the background density. The driver beam (laser or particles), propagates into a moving window of 48μm($x$)×48μm($y$)×48μm($z$) size and 1200×480×480 cell number with 6 macro-particles in each cell. The laser beam is linearly polarized along the $y$-axis, following a Gaussian profile, $E_L=aw_0^2/w^2(x)\sin^2(\pi t/2\tau_0)\sin(\pi r^2/\lambda R)\exp[-r^2/w^2(x)]$, with normalized peak amplitude $a=eE/m\omega c$, $r^2=y^2+z^2$, wavelength $\lambda=800nm$, $w(x)=w_0\{[(x-x_p)^2+x_R^2]/x_R^2\}^{1/2}$, $R=(x^2+x_R^2)/x$, Rayleigh length $x_R=\pi w_0^2/\lambda$, width $w_0=10\lambda(8\mu m)$ and duration $\tau_0=10\lambda/c=26.7fs$, respectively. The electron driver beam also takes a Gaussian profile [8] $n_b(r,\xi=x-ct)=n_{b0}\exp(-r^2/2\sigma_r^2-\xi^2/2\sigma_l^2)$, where $\sigma_l$, $\sigma_r$ are the longitudinal and transverse beam sizes and $n_{b0}=1.5\times10^{19}cm^{-3}$ denotes the peak density of the driving beam, respectively. The simulation time step is $\Delta t=0.01\lambda/c$ ensuring that the largest precession angle in each time step $|\theta_{s,max}|=\Omega_{max}\Delta t\sim0.02\pi B_{max}$ ($B$ is normalized by $m\omega/e$) is sufficiently small ($|\theta_{s,max}|<<2\pi$).

### III. RESULTS

The electron density distributions of a typical LWFA and the corresponding transverse fields based on density down ramp injection are displayed in Fig. 1(a). The results are collected for $a=2.5$, $\alpha=4$ and $n_0=10^{18}cm^{-3}$. A sphere-like bubble is generated by the laser ponderomotive force. The injected electrons are located at the rear of the bubble and continuously gain energy from the longitudinal field. The total charge of the injected beam is about 62nC and peak current reaches 9.2kA. Seen from the directional arrows in the transverse $y$-$z$ plane, we notice that the transverse fields satisfy $\boldsymbol{B}_T\sim-B_\phi\boldsymbol{e}_\phi$ and $\boldsymbol{E}_T\sim E_r\boldsymbol{e}_r$. Such field distributions indicate a central force on the injected electrons. We show the trajectories of electrons initially located at a radius of $|r_i|=6.4\mu m$ in Fig. 1(b) and mark their spin orientations during the injection phase for selected electrons at distinct coordinate angles $\phi=\tan^{-1}(z/y)$ in the range $-\pi/2$ to $\pi/2$ (since $\phi+\pi$ and $\phi$ correspond to the same axis with opposite directions, it suffices to treat the region $\phi\in[-\pi/2, \pi/2]$). It is seen that due to the cylinder symmetry of the bubble forces, all electrons are focused to the center, following almost identical trajectories about the central axis. However, the spin precessions are different for electrons of varying $\phi$. In particular, the electron spin is preserved at $\phi=0$ (along the $y$-axis) during injection while changes dramatically for $\phi=\pm\pi/2$ (along the $z$-axis).

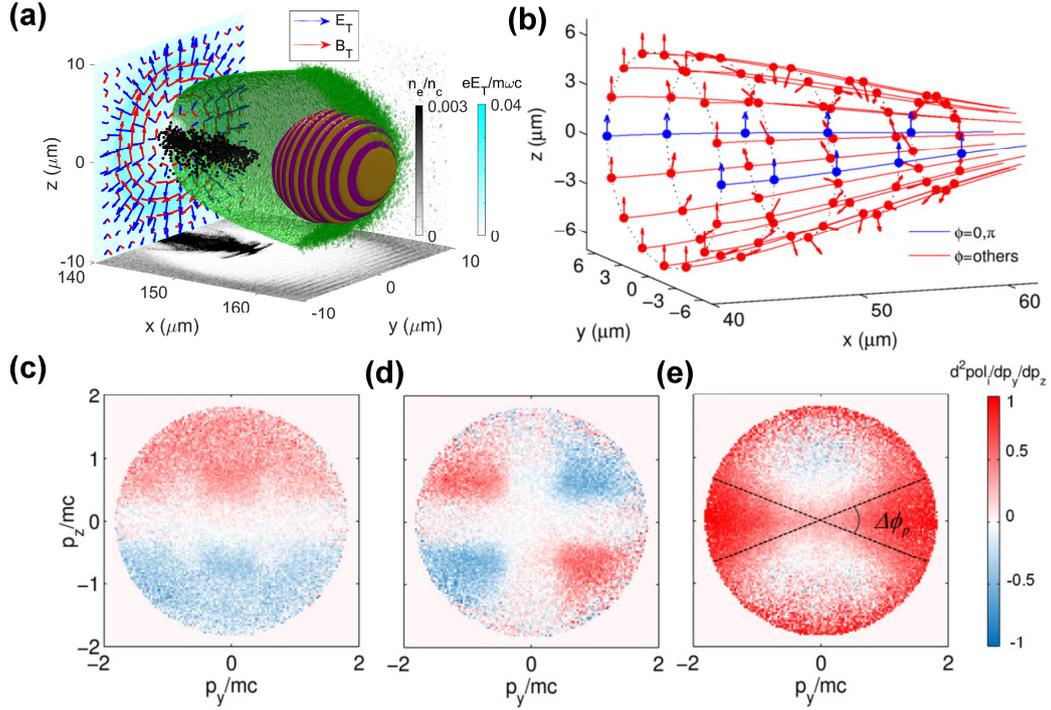

**Figure 1**. (a) A typical LWFA structure and transverse field distributions, with the bubble (green surface), the laser beam (yellow-purple iso-surfaces for $|E_y|=1.5$), the injected electrons (black dots), the electron density (the $x$–$y$ plane) and the transverse $E$-field ($E_T$, blue arrows) and $B$-field ($B_T$, red arrows). The amplitude of $E_T=(E_y^2+E_z^2)^{1/2}$ ($x=150\mu m$ plane) is projected onto the $x=140\mu m$ plane. (b) Electron trajectories (solid lines) and spin orientations (arrows) during injection with $\phi \in [-\pi/2, \pi/2]$ at injection radius $|r_i|=6.4\mu m$, with $\phi=0$ marked in blue and the rest in red. The polarization (average spin components in mesh-grids) distributions in the transverse phase space ($p_y$-$p_z$) (c, d and e respectively) along the $x$, $y$ and $z$ axis at 800fs. The results are collected for simulation parameters of $a=2.5$, $\alpha=4$, $n_0=10^{18}$cm$^{-3}$ and initial electron spins are set along the $z$-axis. The cross-shaped area defined by the phase angle $\Delta\phi_p$ in (e) represents the high polarization region.

We note that the trajectories basically remain in a plane defined by the coordinates and the central axis for each electron. It indicates that the azimuthal angle is the same in both the coordinate space ($\phi$) and the transverse phase space (defined as $\phi_p=\tan^{-1}(P_z/P_y)$) during the injection. For instance, the electrons located at $\phi$ in the range of $[-\pi/2, \pi/2]$ would be directed along $\phi_p \sim [-\pi/2, \pi/2]$ in the transverse directions. Therefore, spins should be preserved for electrons moving close to the $y$-axis ($|p_z|<<|p_y|$) and change significantly around the $z$-axis ($|p_z|>>|p_y|$). To further prove the statement, we give the polarization $\Sigma s_i/N$ distributions in $p_y$-$p_z$ space along $x$, $y$ and $z$ axis (averaging the corresponding spin components in $p_y$-$p_z$ mesh-grids). As illustrated in Fig. 1(e) for beam polarization along the $+z$-axis (the initial spin orientation), where the polarization purity peaks in the vicinity of $\phi_p \sim 0$ and vanishes otherwise, exhibiting a saddle-like distribution. The polarization along the $y$-axis in Fig. 1(d) is also centrally symmetric but peaks at $\phi_p \sim \pm\pi/4(|p_z|\sim|p_y|)$. The one along the $x$-axis is symmetric about the

$p_z$=0 axis and averaged out at specific $\phi_p$ as shown in Fig. 1(c). Simulations suggest that the peak value of the polarization along the z-axis is well retained at >80% among the cross region of $|\phi_p|<\pi/4$, which allows for spin filtering in the transverse phase space.

To explain the above observations, we consider a quasi-static scenario, where the bubble fields are cylinder symmetric, *i.e.* where **B**~-$B_\phi \mathbf{e}_\phi$, $\beta$~$\beta_x$, $\boldsymbol{\beta}\times\mathbf{E}$~$E_r\beta_x\mathbf{e}_\phi$ [42-44]. Such field distributions are also seen in our simulations (Fig.1 (a)). As already stated in previous works [27-29,45], the spin is preserved during steady acceleration but precesses dramatically during the injection phase. That is why we focus on the spin dynamics during injection, where typically one has $\gamma\ll 1/a_e$ [42,46] and the second term of Eq.(1b) can be neglected. With these simplifications, the precession frequency $\boldsymbol{\Omega}$ and the equation of motion in the transverse direction can be written as:

$$\boldsymbol{\Omega} = -\frac{e}{m}\left(\frac{B_\phi}{\gamma} + \frac{\beta_x}{\gamma+1}\frac{E_r}{c}\right)\mathbf{e}_\phi. \tag{2}$$

$$mc\frac{\partial}{\partial t}(\boldsymbol{P_T}) = e(E_r + c\beta_x B_\phi)\mathbf{e_r} = F_r \mathbf{e_r} \tag{3}$$

An obvious effect revealed by Eq. (3) is that the electrons only feel radial forces in transverse directions $\partial \boldsymbol{P_T}/\partial t$~$\partial \boldsymbol{P_r}/\partial t$. Since electrons injected at an azimuthal angle $\phi$ carry zero initial azimuthal momentum, *i.e.*, $\boldsymbol{P_\phi}$~0, the azimuthal angles are equivalent in the coordinate space and the phase space $\phi$~$\phi_p$ during the whole injection process, which agrees with the trajectories in Fig. 1(b). The distributions of injected electrons with their transverse momentum vectors in Fig. 2(a) further demonstrate that momentum vectors are almost parallel to the coordinate vector direction, with negligible minor discrepancies. The difference between the two defined by $\phi$–$\phi_p$ is counted over all injected electrons, which shows a root-mean square (RMS) value of only $0.063\pi$ in Fig. 2(b).

Known from Eq. (3), the coordinate angle $\phi$ of an electron is always fixed (there is no lateral force perpendicular to $\mathbf{e_r}$) which guarantees that the precession axis is stationary (along $\mathbf{e}_\phi$). The precession angle satisfies $\Delta\theta_s$=<$\boldsymbol{\Omega}$>$\Delta T$ where $\Delta T$~$4|r_i|/c$ is the injection time and $r_i$ is the injection radius [47-49]. The spin vector after injection (starting at $t$=$t_0$) can then be deduced by $\mathbf{s}(t_0+\Delta T)$=$\mathbf{s}_{//}(t_0+\Delta T)$+$\mathbf{s}_\perp$ $(t_0+\Delta T)$=$\mathbf{s}_{//}(t_0)$+cos($\Delta\theta_s$)$\mathbf{s}_\perp(t_0)$+sin($\Delta\theta_s$)$\mathbf{e}_\phi\times \mathbf{s}_\perp(t_0)$, where $\mathbf{s}_{//}(t_0)$=[$\mathbf{s}(t_0)\cdot\mathbf{e}_\phi]\mathbf{e}_\phi$=cos$\phi \mathbf{e}_\phi$ is the component parallel to $\boldsymbol{\Omega}$ and $\mathbf{s}_\perp(t_0)$= $\mathbf{s}(t_0)$-$\mathbf{s}_{//}(t_0)$=sin$\phi \mathbf{e_r}$ is perpendicular to $\boldsymbol{\Omega}$. One obtains the spin component along each axis following $s_x(t_0+\Delta T)$=$\mathbf{s}(t_0+\Delta T)\cdot \mathbf{e_x}$=-sin($\Delta\theta_s$)sin$\phi$, $s_y(t_0+\Delta T)$=$\mathbf{s}(t_0+\Delta T)\cdot\mathbf{e_y}$=sin2$\phi$[cos($\Delta\theta_s$)–1]/2 and $s_z(t_0+\Delta T)$=$\mathbf{s}(t_0+\Delta T)\cdot \mathbf{e_z}$=cos$^2\phi$+sin$^2\phi$cos($\Delta\theta_s$). These results immediately disclose a key feature in wakefield acceleration: the spin distribution is a function of the coordinate angle $\phi$. In the following, we will show that the dependence holds for the beam polarization.

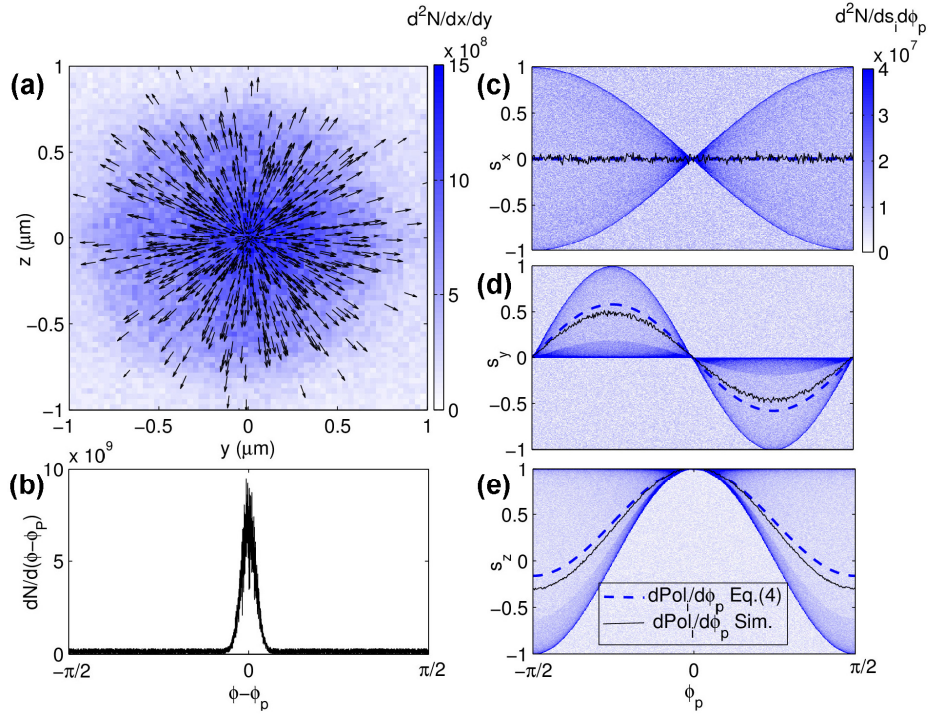

**Figure 2**. The number density in the *y-z* plane (a) and the value of $\phi-\phi_p$ (b) of all injected electrons. The black arrows in (a) denote transverse momentum-vector directions for randomly selected 1000 electrons. The particle densities in the $\phi_p$-$s_i$ space where $i=x$ (c), $y$ (d) and $z$ (e), respectively. The average spin (polarization) of each is shown from the theoretical analysis (blue-dashed) and the simulations (black solid, calculated from integrating the density along the $s_i$ axis). Simulation parameters are the same as those in Fig. 1.

According to Refs. [38,40,50], the average fields in the plasma bubble take the form $E_r \sim cB_\phi \sim en(x)r/4\varepsilon_0$. Taking $\langle\beta_x\rangle \sim 1/2$, $\gamma \sim 1$, $\langle n(x)\rangle \sim (\alpha+1)n_0/2$ and $\langle r\rangle \sim r_i/2$ during the injection phase, one obtains $\langle\Omega\rangle \approx -5e\langle B_\phi\rangle/4m \approx -5e^2(\alpha+1)n_0 r_i/64m\varepsilon_0$, thus $\Delta\theta_s \sim \langle\Omega\rangle \sim \Delta T \sim -\eta r_i|r_i|$ with $\eta=5e^2(\alpha+1)n_0/16mc^2\varepsilon_0$. For simplicity, we assume a homogeneous electron distribution within the cylindrical injection volume of $|r_i|\leq r_b(x_p)$ [27-29], where $r_b(x_p)$ is the bubble radius at the density peak, satisfying $r_b^2(x_p) \sim 4\sigma_l\sigma_r(n_{b0}/\alpha n_0)^{1/2}$ for PWFA [42,44] and $r_b^2(x_p) \sim 4amc^2\varepsilon_0/\alpha e^2 n_0$ for LWFA [51]. Therefore, combining with the fact that $\phi \sim \phi_p$, the polarization at certain $\phi_p$ becomes $dPol_i/d\phi_p \sim dPol_i/d\phi = \int_{-r_b(x_p)}^{r_b(x_p)} s_i |r|/r_b^2(x_p) dr$, where the three components ($i=x, y, z$) are:

$$\frac{dPol_x}{d\phi_p} = \frac{1}{r_b^2(x_p)} \int_{-r_b(x_p)}^{r_b(x_p)} \sin(\eta r|r|)\sin\phi_p |r| dr = 0 \quad (4a)$$

$$\frac{dPol_y}{d\phi_p} = \frac{\sin 2\phi_p}{2r_b^2(x_p)} \int_{-r_b(x_p)}^{r_b(x_p)} [\cos(\eta r|r|) - 1]|r| dr = \frac{\sin 2\phi_p}{2}[sinc(\psi) - 1] \quad (4b)$$

$$\frac{dPol_z}{d\phi_p} = \frac{1}{r_b^2(x_p)} \int_{-r_b(x_p)}^{r_b(x_p)} [\cos^2\phi_p + \sin^2\phi_p \cos(\eta r|r|)]|r| dr = 1 - \sin^2\phi_p[1 - sinc(\psi)] \ . \quad (4c)$$

Here $\psi$ is a normalized parameter denoting the largest precession angle under certain parameters, *e.g.*,

$\psi=\eta r_b{}^2=5e^2(\alpha+1)\sigma_l\sigma_r(n_{b0}n_0)^{1/2}/4\alpha^{1/2}mc^2\varepsilon_0$ for PWFA and $\psi=5a(\alpha+1)/4\alpha$ for LWFA. The spin distributions vs. $\phi_p$ for the LWFA case are illustrated in Figs. 2(c), 2(d) and 2(e) ($\psi=3.91$). The theoretical estimation from Eq. (4) shows good agreement with our simulations. Other cases, like PWFA and LWFA driven by circularly polarized light, are also well described by Eq. (4).

Equation (4a) is confirmed by the simulation results in Figs. 1(c) and 2(c), where the polarization is close to 0 along the longitudinal axis (*x*-axis). At any $\phi_p$ the spin values are anti-symmetric around the $s_x=0$ axis. Although the longitudinal component for each spin can be non-zero due to precession, a pair of centrally symmetric electrons precesses around opposite axes at the same frequency, resulting in opposite values of the longitudinal spin component. The polarization along the *x*-axis naturally disappears in the statistic average. With regard to the $s_y$ distribution, as exhibited in both Figs. 1(d) and 2(d), the polarization peaks at $\phi_p=\pm\pi/4$. According to Eq. (4b), both $s_\perp$ and $s_{//}$ contain a factor $\sin(2\phi_p)/2$ that contributes to the *y* component and the maximum absolute value $(1-\text{sinc}(\psi))$ is achieved at $\pi/4$.

The highest degree of polarization is preserved in the initial spin orientation. Equation (4c) clearly shows that the polarization along the *z*-axis increases when $|\phi_p|$ is approaching zero, due to the $-\sin^2\phi_p$ term contributed by $s_\perp$ and $s_{//}$. In particular, the polarization reaches 100% at $|\phi_p|\sim0$ as illustrated in Fig. 2(e). This remarkable feature suggests an efficient method to maximize the beam polarization by selecting electrons around the phase-space angle $\phi_p$.

We therefore propose the spin filter approach to obtain electron bunches of high polarization purity. As sketched in Fig. 3(a), the driver beam excites a plasma wakefield in the pre-polarized plasma target, which accelerates the injected electrons to high energies. After the acceleration, the electron beam can be filtered by an X-shaped slit. It should be mentioned that the spin distribution in the phase space is maintained during the long propagation distance after acceleration, because the self-generated field of the electron beam still follows the form in Eq.(3). The slit is placed perpendicular to *x* with an opening angle $a_s$, so that electrons with $\phi_p$ satisfying $-\Delta\phi_p/2\leq\phi_p\leq\Delta\phi_p/2$ ($\Delta\phi_p=\pi-a_s$) can pass and all other electrons are blocked. If we integrate Eqs. (4a) to (4c) we find the total polarization components $Pol_i(\Delta\phi_p) = \int_{-\Delta\phi_p/2}^{\Delta\phi_p/2} dPol_i /\Delta\phi_p$ within the selected angular range

$$Pol_x(\Delta\phi_p) = Pol_y(\Delta\phi_p) = 0 \qquad (5a)$$

$$Pol_z(\Delta\phi_p) = \frac{1+\text{sinc}(\Delta\phi_p)}{2} + \frac{1-\text{sinc}(\Delta\phi_p)}{2}\text{sinc}(\psi) \qquad . \qquad (5b)$$

Such spin filtering only preserves the polarization in the *z*-direction (the pre-polarized direction). As exhibited in Fig. 3(b), the simulation results for both LWFA and PWFA are in agreement with Eq. (5). The polarization decreases with larger values of $\Delta\phi_p$. To be specific, the polarization is 96% for LWFA and 98% for PWFA at $\Delta\phi_p=\pi/10$ while the polarization without the spin filter ($\Delta\phi_p=\pi$) is only 35% for LWFA and 49% for PWFA.

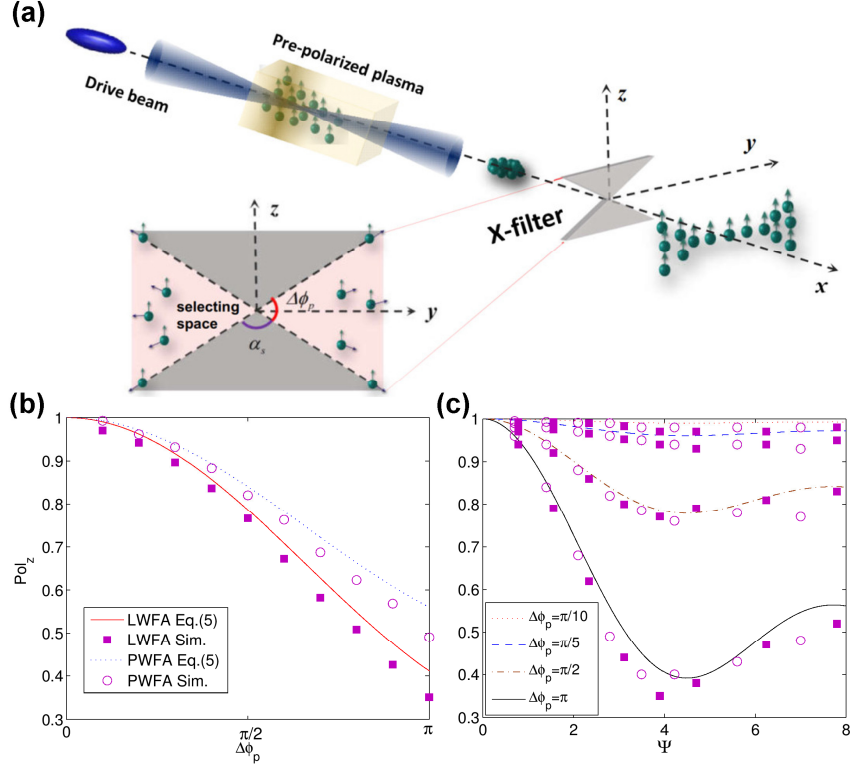

**Figure 3**. (a) Sketch of a spin-filter using the X-shaped slit to produce polarized electron beams, where the slit selects electrons within the region [$-\Delta\phi_p/2$, $\Delta\phi_p/2$]. (b) Beam polarization along the $z$-axis ($Pol_z$) as a function of the selecting angle $\Delta\phi_p$ for LWFA (red solid and cyan square) and PWFA (blue-dotted and cyan circle). The simulation results are collected with the same parameters as in Fig. 2. (c) $Pol_z$ as a function of the normalized parameter $\psi$ from simulations for LWFA (cyan squares) and PWFA (cyan circles) with $\Delta\phi_p=\pi/10, \pi/5, \pi/2, \pi$. The simulations results are given for $a=0.5\sim5$ for LWFA and $\sigma_l=0.8\sim4\mu m$, $\sigma_r=1.6\sim3.2\mu m$, $n_{b0}=1.5\times10^{19}cm^{-3}$ for PWFA with $\alpha=4$ and $n_0=10^{18}cm^{-3}$. The theoretical predictions are show by lines.

The spin-filter mechanism is universal in a large parameter range. The beam polarization as a function of the dimensionless parameter $\psi$ is summarized in Fig. 3(c), for different parameter combinations. While in general smaller selecting angles induce high polarization purity, we see that the beam polarization is already close to 80% for $\Delta\phi_p=\pi/2$ (90%, for $\Delta\phi_p=\pi/5$ and $\pi/10$). In other words, the beam polarization is significantly purified by filtering out only half of the electron flux. Without spin filtering, the beam flux would be strongly restricted to preserve the polarization. As illustrated in Fig. 3(c), it $\psi<2$ is required to maintain a beam polarization above 80% when there is no filtering ($\Delta\phi_p=\pi$). Such restrictions disappear in our X-filter strategy. One can find from Fig. 3(c) that for large value of $\psi$ (representing the beam flux), the beam polarization after spin filtering $\Delta\phi_p \leq \pi/2$ is above 80% throughout the full region.

An interesting phenomenon is that the polarization is not monotone with $\psi$. As a matter of fact, the $z$-component of the spin $s_z=\cos^2\phi+\sin^2\phi\cos(\Delta\theta_s)$ oscillates with $\Delta\theta_s$, decreasing in the range of $[2K\pi, (2K+1)\pi]$ and increasing in the range of $[(2K+1)\pi, (2K+2)\pi]$, where $K$ is the integer. The intrinsic of the spin-filter mechanism is attributed to the field geometry of the wakefield. For transverse initial polarization, the precession axis is along $e_\phi$, the rotation spin component $s_\perp$ is anisotropic for different $\phi_p$ around the procession axis $e_\phi$. As a consequence, the degree of spin precession depends on $\phi_p$, which leads to a disparity of polarization in the transverse phase space. However, for longitudinal initial polarization, the rotation component is $|s_\perp|=1$ for all electrons, which means that the precession degree is independence of $\phi_p$. We conclude that the spin-filter mechanism is not valid for the longitudinal case.

## IV. DISCUSSIONS

The results shown above are based on an ideal situation, where the bubble is in perfect cylindrical symmetry, the plasma is cold initially, the drive beams are precisely aligned to the filter center and the filter is regarded as wireless thin. However, in a real experiment, the bubble is usually not so perfectly symmetric, the plasma gains initially temperature from pre-pulses; the drive beams jitter for high power laser systems and the filter has a finite size. This section is devoted to the robustness of our method against the imperfections.

To evaluate imperfect plasma wakefield effects (field asymmetry), we have introduced two parameters $(\rho_1, \rho_2)$ in the drive beam profile in our 3D PIC simulations:

$$E_L = a[1 + \rho_2\delta(y,z)]w_0^2/w^2(x)\sin^2(\pi t/2\tau_0)\sin(\pi r^2/\lambda R)\exp\{-[\rho_1 y^2 + (2-\rho 1)z^2)]/w^2(x)\} \quad (6)$$

where $0<\rho_1<2$ measures the ellipticity of the transverse intensity profile (perfectly circular for $\rho_1=1$), and $0<\rho_2<1$ represents the degree of field fluctuation (zero fluctuation at $\rho_2=0$) multiplied by random numbers uniformly distributed at transverse coordinate $\delta(y,z)$ among the range [-1,1]. The latter represents a type of quite harsh fluctuation that could possibly happen in experiments.

We first investigate the tolerance of parameter $\rho_1$ ($\rho_2$ is set to 0). As shown in Fig. 4(a) and 4(b), for $\Delta\phi_p=\pi/2$ (50% filter) the angular dependence of spin distribution is well preserved at $\rho_1=0.9$ but not so for $\rho_1=0.8$, where the polarization declines in the filter region due to mixing of the spins in asymmetrical field. From the scanning results depicted in Fig. 2(c), the polarization with $\Delta\phi_p=\pi/2$ is above 70% for

$0.9<\rho_1<1.1$, indicating a tolerable fluctuation within ±10% for ellipticity. It can be further relaxed to ±20% ($0.8<\rho_1<1.2$) when choosing a smaller filter $\Delta\phi_p=\pi/5$. We then show the impacts of field fluctuation by setting $\rho_1=1$ and varying $\rho_2$ from 0 to 0.25. As illustrated in Fig. 4(d), the polarization declines due to the uneven distribution of the bubble field. Nevertheless, the polarization still surpasses 70% for $\rho_2<0.1$ when $\Delta\phi_p=\pi/2$ and $\rho_2<0.25$ when $\Delta\phi_p=\pi/5$. These results imply that about 20% fluctuations of laser field amplitude is tolerable for the filter mechanism.

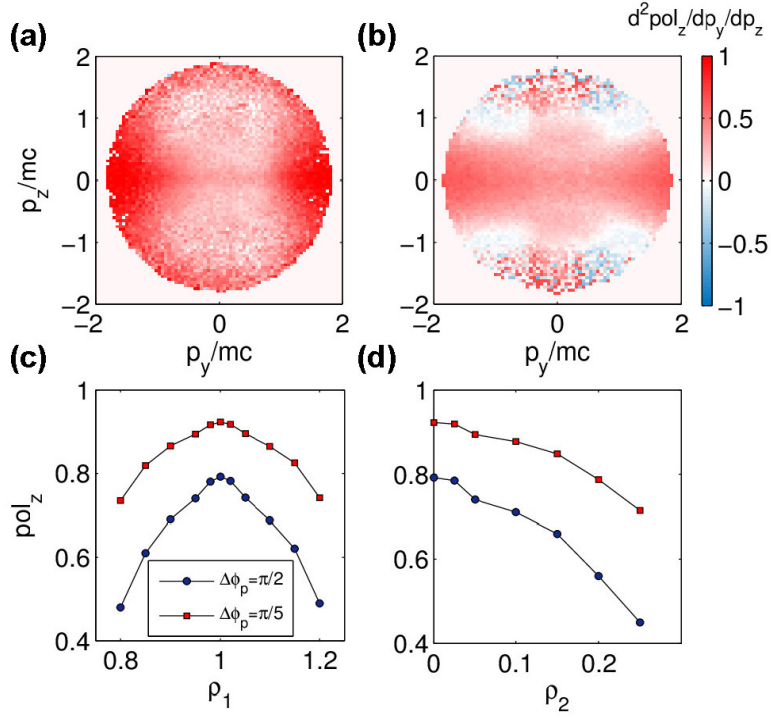

**Figure 4**. The polarization (statistics along z axis) distributions in the transverse phase space ($p_y$-$p_z$) for (a) $\rho_1=0.9$, $\rho_2=0$ and (b) $\rho_1=0.8$, $\rho_2=0$ respectively. The polarization along the z-axis as a function of (c) $\rho_1$ and (d) $\rho_2$ for the selecting angles $\Delta\phi_p=\pi/2$ and $\pi/5$. Other simulation parameters are the same as those in Fig. 1.

As mentioned above, the plasma is not completely cold before arrival of the main pulse. Since the thermal motion of electrons affect the spin filter process through the mapping from $\phi$ and $\phi_p$, we carry out simulations with different initial temperatures (assuming the electrons satisfy the Maxwell-Boltzmann distribution) under the same parameters as in Fig. 1. As illustrated in Fig. 5(a), the polarization is well above 70% for temperatures approaching 1keV. This is fulfilled in typical wakefield acceleration experiments.

In Sec. III, the filter is considered as an ideal plane without thickness. In reality, both transverse size and longitudinal thickness of our designed filter may affect the selecting polarization. These effects canbe seen from Monte Carlo simulations, where the target is set 50cm away from the filter. When the transverse size $L_{ty}$ of the filter increases, the polarization of selecting beam is enhanced while the total charge declines, as shown in Fig. 5(b). The polarization increases to about 90% when $L_{ty}$ reaches 1cm, whereas the charge of the selecting beam drops to about 8pC. The filter thickness $L_{tx}$ has a joint effect with the beam jitter on the selecting process. In simulations, we artificially introduce a position shift (along *y*- and *z*- axis) between the drive beam axis and the X-filter center. The thickness of the filter is also varied. The statistics are shown in Fig. 5(c) and (d). The jitter$_z$ and jitter$_y$ represent the beam center position with respect to (*y*,*z*)=(0,0) to the front of the filter. Not surprisingly, one finds the polarization decline with the augment of both jitter and longitudinal thickness. For thicker filters, the tolerant range of jitter is smaller. However, the polarization is well beyond 70% if the position jitter<1mm and $L_{tx}$<4cm. These results provide guidance for future experiments.

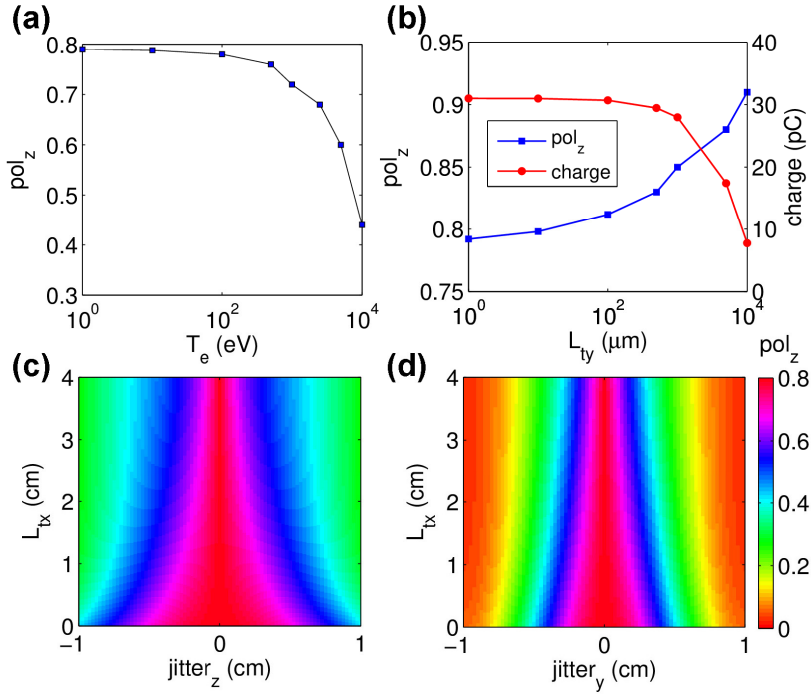

**Figure 5**. (a) The polarization along *z* axis as a function of initial electron temperature. (b) The polarization (blue squared) and selecting charge (red circled) as a function of the transverse size (along the y-axis) of the filter $L_{ty}$. The polarization as a function of longitudinal thickness (along *x* axis) $L_{tx}$ and drive beam jitter in *z* direction (*jitter$_y$*=0) (c) and in *y* direction (*jitter$_z$*=0) (d). The target is set 50cm away from the filter. The selecting angles $\Delta\phi_p = \pi/2$ while other simulation parameters are the same as those in Fig. 1.

## V. SUMMARY

In conclusion, we investigated the azimuthal dependence of beam polarization in the wakefield and propose a spin-filter strategy for transversely polarized electron beams. By means of PIC simulations and theoretical analyses in the scope of a quasi-static model, we find out that the precession degree varies for certain $\phi_p$ (the azimuthal angle in the phase space). In particular, electron spin directions are almost preserved during acceleration along the $\phi_p=0$ axis ($\phi_p=\pi/2$ if initially polarized along y axis) since the rotation components are vanishing. Therefore, applying this effect can maximize the beam polarization via selection of electrons in certain phase-space regions, at moderate cost of the beam flux. According to the theoretical model and 3D simulations, about 80% beam polarization can be obtained by filtering out half of all the electrons, compared to about 35% for the unfiltered beam. The robust of the method is discussed against various imperfect situations, which will guide future experiments in high power laser facilities. The filtering mechanism is universal in a large parameter range. The limitations for the driver-beam parameters and the total beam flux are relieved, which hamper previous schemes. These highly polarized electron beams are advantageous in applications such as future electron-positron colliders.

## ACKNOWLEDGEMENTS

This work is supported by the Strategic Priority Research Program of Chinese Academy of Sciences (Grant No. XDB 16010000), the National Science Foundation of China (Nos. 11875307, 11935008) and the Recruitment Program for Young Professionals. MB and AH acknowledge support through the HGF-ATHENA project.

*jill@siom.ac.cn
#bfshen@mail.shcnc.ac.cn
†ruxinli@mail.siom.ac.cn